\documentstyle[12pt]{article}
\catcode`\@=11

\newcommand{\be}{\begin{equation}}
\newcommand{\ee}{\end{equation}}
\newcommand{\ba}{\begin{eqnarray}}
\newcommand{\ea}{\end{eqnarray}}

\def\12{\frac{1}{2}}
\def\fr{\frac}
\def\pr{\partial}
\def\prd{\partial \cdot}
\def\btd{\bigtriangledown}

\def\a{\alpha}
\def\b{\beta}

\def\G{\Gamma}

\def\d{\delta}
\def\e{\epsilon}

\def\h{\eta}

\def\L{\Lambda}
\def\m\mu 
\def\n{\nu}

\def\f{\phi}

\def\vf{\varphi}

\def\dsll{\not {\! \pr}}
\def\psisl{\not {\! \! \psi}}
\def\esl{\not {\! \epsilon}}
\def\ssl{\not {\! \cal S}}
\def\@normalsize{\@setsize\normalsize{15pt}\xiipt\@xiipt
\abovedisplayskip 14pt plus3pt minus3pt%
\belowdisplayskip \abovedisplayskip
\abovedisplayshortskip  \z@ plus3pt%
\belowdisplayshortskip  7pt plus3.5pt minus0pt}
\def\small{\@setsize\small{13.6pt}\xipt\@xipt
\abovedisplayskip 13pt plus3pt minus3pt%
\belowdisplayskip \abovedisplayskip
\abovedisplayshortskip  \z@ plus3pt%
\belowdisplayshortskip  7pt plus3.5pt minus0pt
\def\@listi{\parsep 4.5pt plus 2pt minus 1pt
            \itemsep \parsep
            \topsep 9pt plus 3pt minus 3pt}}

\def\underline#1{\relax\ifmmode\@@underline#1\else
        $\@@underline{\hbox{#1}}$\relax\fi}
\@twosidetrue
\relax

\catcode`@=12

\catcode`\@=11

\def\section{\@startsection{section}{1}{\z@}{3.5ex plus 1ex minus
   .2ex}{2.3ex plus .2ex}{\large\bf}}

\def\thesubsection{\Roman{section}-\arabic{subsection}}

\def\ps@headings{\def\@oddfoot{}\def\@evenfoot{}
\def\@oddhead{\hbox{}\hfill
        \makebox[.5\textwidth]{\raggedright\ignorespaces --\thepage{}--
        \hfill }}
\def\@evenhead{\@oddhead}
\def\subsectionmark##1{\markboth{##1}{}} }
\renewcommand{\subsection}[1]{\addtocounter{subsection}{1}
\vspace{2.5mm}\par\noindent {\em \thesubsection . #1}\par
 \vspace{0.5mm} }

\ps@headings

\catcode`\@=12

\relax

\def\figcap{\section*{Figure Captions\markboth
        {FIGURECAPTIONS}{FIGURECAPTIONS}}\list
        {Fig. \arabic{enumi}:\hfill}{\settowidth\labelwidth{Fig. 999:}
        \leftmargin\labelwidth
        \advance\leftmargin\labelsep\usecounter{enumi}}}
 \relax
\def\tablecap{\section*{Table Captions\markboth
        {TABLECAPTIONS}{TABLECAPTIONS}}\list
        {Table \arabic{enumi}:\hfill}{\settowidth\labelwidth{Table
999:}
        \leftmargin\labelwidth
        \advance\leftmargin\labelsep\usecounter{enumi}}}
 \relax
\def\reflist{\section*{References\markboth
        {REFLIST}{REFLIST}}\list
        {[\arabic{enumi}]\hfill}{\settowidth\labelwidth{[999]}
        \leftmargin\labelwidth
        \advance\leftmargin\labelsep\usecounter{enumi}}}
 \relax

\catcode`\@=11

\def\marginnote#1{}
\newcount\hour
\newcount\minute
\newtoks\amorpm
\hour=\time\divide\hour by60
\minute=\time{\multiply\hour by60 \global\advance\minute by-
\hour}
\edef\standardtime{{\ifnum\hour<12 \global\amorpm={am}%
    \else\global\amorpm={pm}\advance\hour by-12 \fi
    \ifnum\hour=0 \hour=12 \fi
    \number\hour:\ifnum\minute<100\fi\number\minute\the\amorpm}}
\edef\militarytime{\number\hour:\ifnum\minute<100\fi\number\minute}
\def\draftlabel#1{{\@bsphack\if@filesw {\let\thepage\relax
  \xdef\@gtempa{\write\@auxout{\string
    \newlabel{#1}{{\@currentlabel}{\thepage}}}}}\@gtempa
    \if@nobreak \ifvmode\nobreak\fi\fi\fi\@esphack}
     \gdef\@eqnlabel{#1}}
\def\@eqnlabel{}
\def\@vacuum{}
\def\draftmarginnote#1{\marginpar{\raggedright\scriptsize\tt#1}}
\def\draft{\oddsidemargin -.5truein
        \def\@oddfoot{\sl preliminary draft \hfil
        \rm\thepage\hfil\sl\today\quad\militarytime}
        \let\@evenfoot\@oddfoot \overfullrule 3pt
        \let\label=\draftlabel
        \let\marginnote=\draftmarginnote
   
\def\@eqnnum{(\theequation)\rlap{\kern\marginparsep\tt\@eqnlabel}%
\global\let\@eqnlabel\@vacuum}  }
\def\preprint{\twocolumn\sloppy\flushbottom\parindent 1em
        \leftmargini 2em\leftmarginv .5em\leftmarginvi .5em
        \oddsidemargin -.5in    \evensidemargin -.5in
        \columnsep 15mm \footheight 0pt
        \textwidth 250mmin      \topmargin  -.4in
        \headheight 12pt \topskip .4in
        \textheight 175mm
        \footskip 0pt
        
\def\@oddhead{\thepage\hfil\addtocounter{page}{1}\thepage}
        \let\@evenhead\@oddhead \def\@oddfoot{} \def\@evenfoot{}  }
\def\titlepage{\@restonecolfalse\if@twocolumn\@restonecoltrue\onecolumn
     \else \newpage \fi \thispagestyle{empty}\c@page\z@
        \def\thefootnote{\fnsymbol{footnote}} }
\def\endtitlepage{\if@restonecol\twocolumn \else  \fi
        \def\thefootnote{\arabic{footnote}}
        \setcounter{footnote}{0}}  
\catcode`@=12
\relax

\def\ps@headings{\def\@oddfoot{}\def\@evenfoot{}
\def\@oddhead{\hbox{}\hfill
        \makebox[.5\textwidth]{\raggedright\ignorespaces --\thepage{}--
        \hfill }}
\def\@evenhead{\@oddhead}
\def\subsectionmark##1{\markboth{##1}{}} }

\ps@headings

\relax

\def\firstpage#1#2#3#4#5#6{
\begin{document}


\begin{titlepage}
\nopagebreak
\title{\begin{flushright}
        \vspace*{-1.2in}
        {\normalsize ROM2F-02/18}
\end{flushright}
\vfill {#3}}
\vskip 12pt
\author{\large #4 \\[1.0cm] #5}
\maketitle
\vskip -9mm     
\nopagebreak 
\vskip 48pt
\begin{abstract} {\noindent #6}
\end{abstract}
\vskip 48pt
\begin{center}
{\sl ( \, June, \ 2002 \, ) }
\end{center}
\vfill
\begin{flushleft}
\rule{16.1cm}{0.2mm}\\
$^{\dagger}${ I.N.F.N. Fellow.}\\[-4mm]
\vskip 18 pt
\end{flushleft}
\thispagestyle{empty}
\end{titlepage}}

\date{}
\firstpage{3118}{IC/95/34} {\Large\bf Free geometric equations
for higher spins}  
{Dario Francia$^{\dagger}$ and Augusto Sagnotti} 
{
\small\sl Dipartimento di Fisica\\ [-3mm]
 \small \sl Universit{\`a} di Roma ``Tor Vergata''\\ [-3mm]
\small\sl INFN,
 Sezione di Roma ``Tor Vergata''\\[-3mm]\small\sl Via della Ricerca
Scientifica 1, 00133 Roma, Italy}  
{We show how allowing non-local terms in the field equations
of symmetric tensors uncovers a neat geometry that naturally generalizes 
the Maxwell and Einstein cases. The end results can 
be related to multiple traces of the generalized Riemann curvatures 
${\cal R}_{\alpha_1 \cdots \alpha_{s}; \beta_1 \cdots \beta_{s}}$
introduced by de Wit and Freedman, divided by 
suitable powers of the D'Alembertian operator $\Box$. The conventional
local equations can be recovered by a partial gauge fixing
involving the trace of the gauge parameters 
$\Lambda_{\alpha_1 \cdots \alpha_{s-1}}$, absent in the Fronsdal
formulation. The same geometry underlies the fermionic equations,
that, for all spins $s+1/2$, can be linked via the 
operator $\frac{\not \hskip 1pt \pr}{\Box}$ to those of the spin-$s$ bosons.}
\break
\section{Introduction and summary}

String Theory has long been held by several authors to correspond to a broken 
phase of a higher-spin gauge theory, a viewpoint clearly suggested,
for instance,
by the BRST formulation of free String Field Theory, that encodes
infinitely many higher-spin symmetries in the Stueckelberg mode \cite{sft}. 
However, String Theory presents some clear simplifications with respect to 
unbroken higher-spin theories, well reflected
in the familiar option of associating to large-scale 
phenomena a low-spin low-energy effective description. This is
a general feature of spontaneously broken gauge theories, quite
familiar from simpler examples: for instance,
differently from Q.C.D., at low energies the electro-weak theory
reduces to a low-spin theory with a local   
Fermi coupling, that for many years has been at the heart of weak
interaction phenomenology.  On the other hand, it is in Q.C.D. that
gauge theory comes to full power, with remarkable infrared phenomena
responsible for quark confinement. Even more striking dynamics can thus
be expected from these complicated systems, and this is by itself
an important motivation to try to gain some familiarity with them.

Free covariant equations for fully symmetric tensors
and tensor-spinors were first constructed in
the late seventies by Fronsdal \cite{fron} and Fang and Fronsdal
\cite{ffron}, starting from the massive equations of Singh and 
Hagen \cite{shag}. These are interesting classes of 
higher-spin gauge fields, that in four dimensions exhaust 
all available possibilities, up to dualities, and have the clear
advantage of allowing rather simple unified descriptions. 
Following an important observation of the G\"oteborg
group \cite{goteborg}, that showed how a proper
cubic flat-space vertex could be found for higher spins,
Fradkin and Vasiliev \cite{fvas} have led for many years the search 
for an extension of the free equations to consistent interacting gauge 
theories of higher spins. Arguments related to the gauge
algebra imply that these are bound to involve infinitely many gauge fields
of increasing spins, and in the early nineties Vasiliev finally
arrived at closed-form dynamical equations for symmetric tensors  
$\phi_{\mu_1 \cdots \mu_s}$  of arbitrary rank in mutual interaction 
\cite{vaseq}, but an action principle is still lacking for 
this complicated system.
A crucial input in the constructions of \cite{fvas,vaseq} 
was the inclusion of a cosmological term, that allowed to 
cancel recursively contributions generated by higher-spin gauge
transformations depending on the space-time Weyl tensor, thus
bypassing the difficulties met in earlier attempts \cite{difficult}. 
Various aspects of the work of Vasiliev and collaborators are reviewed
in \cite{revhs}, while recent, related work is described in \cite{recent}.

A peculiar feature of the Fang-Fronsdal
equations is the need for unusual constraints, so that, for instance, 
the bosonic gauge parameters are to be traceless, while
the corresponding gauge fields are to be doubly traceless.
These constraints manifest themselves as symmetry conditions in 
the spinor formalism of \cite{fvas,vaseq}, but appear
less natural in the usual component notation \footnote{
The double trace condition, however, can be related to the 
${\rm OSp}(D-1,1|2)$ structure of the corresponding system with ghosts.
We are grateful to W. Siegel for calling this fact to our attention.}.
This letter is thus devoted to showing how one can formulate the dynamics of 
symmetric tensors and tensor-spinors while foregoing the restrictions
implicit in the Fang-Fronsdal equations. One can well work in generic 
space-time
dimensions, with the proviso that for  $d > 4$ these fields do not 
exhaust all available possibilities. The end result
is rather amusing, since {\it the free equations
contain non-local terms} whenever the gauge fields have more than a pair of
symmetric Lorentz indices, {\it i.e.} in all cases beyond the familiar Maxwell
and Einstein examples. However, all non-local terms can
be eliminated by a partial gauge fixing using the trace (or, for 
fermions, the $\gamma$-trace) of the gauge parameter, that
reduces the geometric equations to the Fang-Fronsdal form. This analysis
will bring us naturally to consider, following de Wit and Freedman
\cite{dewf}, higher-spin
generalizations of the Christoffel connection, ${\Gamma}_{\alpha_1 
\cdots \alpha_{s-1};\beta_1 \cdots \beta_{s}}$, and of the Riemann
curvature, ${\cal R}_{\alpha_1 \cdots \alpha_{s};\beta_1 \cdots 
\beta_{s}}$, that are totally symmetric under the interchange of
any pair of indices within the two sets. In terms of these quantities,
the gauge invariant bosonic field equations will be
\be
\frac{1}{\Box^{p}} \ \prd \
{\cal R}^{[p]}{}_{; 
\alpha_1 \cdots \alpha_{2p+1}} \  =\  0
\label{oddcurv}
\ee
for odd spins $s=2p+1$, and
\be
\frac{1}{\Box^{p-1}} \
{\cal R}^{[p]}{}_{;
\alpha_1 \cdots \alpha_{2p}} \ =\ 0
\label{evencurv}
\ee
for even spins $s=2p$. Here and in the following, a superscript
$[p]$ denotes a $p$-fold trace, while $\prd$ denotes a 
divergence, but for the sake of brevity low-order traces
will be occasionally denoted by ``primes''. Moreover,
we shall work throughout with a ``mostly positive'' Minkowski
metric.

The analogy with the Maxwell and Einstein cases should be evident, and it
is rather pleasing to see a simple pattern
extending to all higher-rank tensors. Let us stress that all these 
equations are manifestly invariant under gauge transformations
without any constraints on the gauge fields or on the corresponding
gauge parameters and that, after a partial gauge fixing, they can all 
be reduced to the conventional, local, Fronsdal form.

This geometric form also results in fermionic equations that are 
closely related to the bosonic
ones. In general, the spin-$(s+1/2)$ fermionic equations can be formally
recovered from the spin-$s$ bosonic operators, properly multiplied by 
$\frac{\not {\hskip 1pt \pr}}{\Box}$, and therefore the 
geometry underlying the
bosonic equations plays a similar, albeit more indirect, role also in the
fermionic case. It is amusing to illustrate right
away this fact, obvious for the Dirac equation, 
for a less evident case, the Rarita-Schwinger equation for spin 3/2, that is
quite familiar from supergravity. This is usually written in the form
\be
\, \gamma^{\mu\nu\rho}\, \partial_\nu \, \psi_\rho \ = \ 0 \ ,
\ee
but once combined with its $\gamma$-trace, it becomes
\be
\, \dsll \, \psi_\mu \ - \ \partial_\mu \psisl \ = \ 0 \ .
\label{rschw}
\ee
The connection with the Maxwell equation that we are advertising can be
exhibited combining it again with its $\gamma$ trace, now multiplied with
$\frac{\pr_\mu {\not \hskip 1pt}\pr}{\Box}$, and the end result is indeed
\be
\frac{\dsll}{\Box} \, \left( \, \Box \psi_\mu \ - \ \partial_\mu \, 
\partial \cdot 
\psi \, \right) \ = \ 0 \ . \label{nlrarita}
\ee

In section 2 we begin by examining the field equation
for spin 3, and we show how to extend the Fronsdal formulation to
fully gauge-invariant, albeit non-local, forms, and how to relate the
latter to local forms involving Stueckelberg fields with higher-derivative 
terms.  In section 3 we
show how one can define via an iterative procedure kinetic operators
for all higher spins, derive their Bianchi identities 
and, making direct 
use of them, construct corresponding Einstein-like tensors. In section 4 we 
recover these equations from the geometric notions of connection and curvature
for higher-spin gauge fields,
originally introduced by de Wit and Freedman  \cite{dewf}. 
While in \cite{dewf} the authors linked the local Fang-Fronsdal equations
to traces of one and two-derivative connections, the full geometric equations
presented here are recovered if one insists on resorting to the 
connection ${\Gamma}_{\alpha_1 \cdots \alpha_{s-1};\beta_1 \cdots \beta_{s}}$
and to the corresponding curvature, that for a spin-$s$ field
contain, respectively, $s-1$ and $s$ derivatives. Whereas unconventional,
these are natural ingredients of higher-spin kinetic operators, that
in general should contain both the
D'Alembertian operator $\Box$ and additional terms with up to
$s$ free derivatives. Hence, the non-local structure exposed here 
is unavoidable in our fully covariant setting.
In addition, it anticipates similar properties of the higher-spin
interactions. It is conceivable, although by no means clear to the
authors at the time of this writing, that corresponding simplifications
could take place if the equations of \cite{vaseq} were
formulated along these lines. 
A related observation is that the BRST charge of world-sheet 
reparametrizations, that lies at the heart of String Field Theory,
embodies a massive dynamics of the Fronsdal type, some aspects of which 
are manifest in the constructions of \cite{buch}, that therefore bear a 
direct relationship to the present work, although the field equations 
are presented there in a local form with compensators that does not exhibit
their link with the curvatures.
\section{The spin-3 case}

Let us begin by describing the spin-3 Fronsdal equation \cite{fron}, 
that for the sake of brevity we shall write in the form
\be
{\cal F}_{123} \ = \ 0 \ ,  \label{fr1}
\ee
where
\be
{\cal F}_{123} \equiv \Box \phi_{123} - (\pr_{1} \prd \phi_{23} + 
\pr_{2} \prd \phi_{13} + \pr_{3} \prd \phi_{12}) + 
\pr_{1}\pr_{2}\phi_{3}^{\;'}
+ \pr_{1}\pr_{3}\phi_{2}^{\;'} + \pr_{2}\pr_{3}\phi_{1}^{\;'} \ ,
 \label{frspin3}
\ee
exposing only the subscripts of the three Lorentz indices involved.
A gauge transformation of the spin-3 field $\phi_{123}$,
\be
\delta \phi_{123} \ = \ \partial_1 \Lambda_{23}\ + \ \partial_2 \Lambda_{31}
\ + \
\partial_3 \Lambda_{12} \ , \label{deltaphi3}
\ee
transforms ${\cal F}$ according to
\be
\delta {\cal F}_{123} \ = \ 3 \, \partial_1  \partial_2  \partial_3 \; \Lambda\;' \ ,
\ee
and therefore, as is well known, ${\cal F}$ is gauge invariant only if the
parameter is subject to the constraint
\be
\Lambda{\;'} \ = \ 0 \ . \label{fronsdaltraceless}
\ee

An additional subtlety, already met in the spin-2 case, is that (\ref{fr1})
does not follow directly 
from a Lagrangian. In order to proceed, 
one must therefore introduce an analogue of the linearized Einstein tensor,
\be
{\cal G}_{123} \ = \ {\cal F}_{123} \ - \ \frac{1}{2} \left( \,
\eta_{12} {\cal F}\;'_3 +
 \eta_{23} {\cal F}\;'_1 + \eta_{31} {\cal F}\;'_2 \, \right) \ ,
\ee
where $\h$ denotes the Minkowski metric.
The Bianchi identity
\be
\partial \cdot {\cal F}_{23} \ = \ \frac{1}{2} \left(\, 
\partial_2  {\cal F}\;'_{3} \, +\, 
 \partial_3  {\cal F}\;'_{2} \, \right) \ ,\label{bianchi3}
\ee
then implies that
\be
\partial \cdot {\cal G}_{23} \ = \ - \ \frac{1}{2} \ \eta_{23} \ 
\partial \cdot {\cal F}\;' \ ,
\label{bianchieinstein3}
\ee
and together with eq. (\ref{fronsdaltraceless}) this result 
is instrumental in deriving a gauge-invariant Lagrangian for this system,
since
\be
\partial \cdot {\cal F}\;' \ = \ 3\, \Box \, \prd \phi\;' \ - \ 2\, 
\prd \prd \prd \phi
\ee
does not vanish identically. Integrating
\be
\d {\cal L} \ = \ \d  \phi^{123} \ {\cal F}_{123}
\ee
one can finally recover the Fronsdal action
\be
{\cal L} \ = \ - \, \fr{1}{2}\, (\pr_{\mu}\, \phi_{123})^{2} \, +\,
 \fr{3}{2}\, (\prd \phi_{12})^{2} \, + \, \fr{3}{2}\, (\pr_{\mu}\, \phi_{1}^{\;'})^{2} 
\, + \, \fr{3}{4}\, (\prd \phi^{\;'})^{2} \, + \, 3\, \phi_{1}^{\;'}\, 
\prd \prd \phi^{1} \ .
\label{fronsdal3}
\ee

Our aim is now to extend the gauge symmetry, modifying the kinetic
operator ${\cal F}_{123}$, by itself a sort of 
connection for the trace of the original gauge parameter. This
case is simple enough to arrive quickly at a fully gauge invariant
equation, for instance
\be
{\cal F}_{123} \ - \ 
\frac{1}{3 \; \Box} \, \left(\, \partial_1 \, \partial_2 \, {\cal F}\;'_3
\ + \ \partial_2 \, \partial_3 \, {\cal F}\;'_1 \ + 
\ \partial_3 \, \partial_1 \, {\cal F}\;'_2
\, \right) \ = \ 0 \ . \label{spin3box}
\ee
As in the Fronsdal case, one can then define an Einstein-like tensor
${\cal G}_{123}$ and arrive at
\ba
{\cal{L}} &=& - \, \12 \ (\pr_{\mu}\f_{123})^{2} \ + \ \fr{3}{2}\, 
(\prd \f_{12})^{2} \ + \ \f^{1\;'}\, \prd \prd \f_1 \ - \ \12\, (\prd \f^{\;'})^{2} \nonumber \\ 
                   &&+ \  \12(\pr_{\mu}\f_1^{\;'})^{2} \ + \
\prd \prd \f^1 \ \fr{1}{\Box} \ 
\prd \prd \f_1 \ + \ \prd \prd \prd \f \ \fr{1}{\Box} \ \prd \f^{\;'} \ .
\ea

It is also possible to recast the Lagrangian in a local form, introducing 
a Stueckelberg field $\vf$, such that
\be
\delta \vf \ = \ \Lambda\;' \ ,
\ee
but, as we shall see, the non-local forms will turn out to
underlie an interesting structure. At any rate,
this compensator allows one to construct the two gauge invariant expressions
\ba
&& \pr_\mu \, 
\varphi \ - \ \phi\;'_\mu \ + \ \frac{1}{\Box}\ \prd \prd \phi_\mu
\  - \ \frac{1}{3}\ \frac{\pr_\mu}{\Box^2} \ \prd \prd \prd \phi \ ,
 \\
&& \Box \, \vf \ + \ \frac{2}{3} \ \frac{1}{\Box} \ \prd \prd \prd \phi  \ - \ 
\prd \phi\;' \ , 
\label{scalarcomp}
\ea
and adding suitable combinations of these to ${\cal L}$ 
finally yields the local Lagrangian
\ba
{\cal{L}} &=&- \ \12 \ (\pr_{\mu}\, \f_{123})^{2} \ + \ 
\fr{3}{2}\ (\prd \f_{12})^{2} \ + \
3 \ \f_1^{ \;'}\, \prd \prd \f^{\; 1} \ + \ 
\frac{3}{4}\ (\prd \f^{\;'})^{2}   \nonumber
\\ &&+ \  \frac{3}{2} \ (\pr_{\mu}\, \f_1^{'})^{2} \
- \ \frac{9}{2} \ \vf \, \Box  \, \prd \phi\;' \ + \ 
3 \ \vf \, \prd \prd \prd \phi 
\ + \ \frac{9}{4} \ \vf \; \Box^2 \; \vf \ , \label{localspin3comp}
\ea
that, differently from (\ref{fronsdal3}), is
invariant under gauge transformations with an {\it unconstrained} parameter.

It is interesting to notice, however, that this fully gauge invariant
equation is not unique, another possibility being
\be
{\cal F}_{123} \ - \ \frac{\partial_1 \partial_2 \partial_3}{\Box^2} \
\partial \cdot {\cal F}\;' \ = \ 0 \ , \label{spin3box2}
\ee
that can actually be obtained combining eq.~(\ref{spin3box}) with its trace.
The corresponding non-local Lagrangian
\ba
{\cal{L}}&=& - \ \12 \, (\pr_{\mu}\, \f_{123})^{2} \, +\, \fr{3}{2}\, 
(\prd \f_{12})^{2} \, +\, 3\, \f_1^{ \;'} \ \prd \prd \f^{\; 1} 
\, - \, \prd \prd \prd \f \ \fr{1}{\Box^{2}} \ \prd \prd \prd \f \nonumber \\
&& + \, \ 3\ \prd \prd \prd \f \ \frac{1}{\Box} \ \prd \f^{\; '} \, +\, 
\fr{3}{2} \ (\pr_{\mu}\f_1^{\; '})^{2} \, -\, \fr{3}{2} (\prd \f^{\;'})^{2} 
\label{nonlocal3box2}
\ea
can also be brought to a local form, making use again of the
compensator $\vf$. The end result, obtained adding to (\ref{nonlocal3box2})
the square of the gauge-invariant expression (\ref{scalarcomp}), is again, 
not surprisingly,
the local Lagrangian (\ref{localspin3comp}). Notice that, under a gauge
transformation with generic parameter $\Lambda_{123}$,
\be
\delta \, \left(\,  \frac{\prd {\cal F}\;'}{\Box^2} \, \right)_\mu \ = \ 3 \; \L{\;'}_\mu \ ,
\ee
and therefore the form (\ref{spin3box2}) of the geometric equation
makes it rather transparent that the trace of the gauge parameter
suffices to bring it to the local Fronsdal form.

\section{Kinetic operators for spin-$s$ bosons}

It is possible to extend the results of the previous section
to symmetric tensors of
arbitrary spin. To this end, it is quite convenient to introduce a shorthand
notation that eliminates the need for explicit indices. A generic spin-$s$
tensor will be denoted simply by $\phi$, while derivatives, divergences 
and traces will be denoted by $\pr \f$, $\prd \f$ and $\f{\;'}$ 
$($or, more generally, $\f^{\; [p]}  )$, 
respectively, with the understanding that 
in all cases the implicit indices are totally symmetrized. With this
proviso, one can see that the somewhat unconventional rules
\ba
&& \left( \pr^{\; p} \ \phi  \right)^{\; \prime} \ = \ \Box \ \pr^{\; p-2} \ 
\phi \ + \ 2 \, \pr^{\; p-1} \  \prd \f \ + \ \pr^{\; p} \ \f{\;'} \\
&& \pr^{\; p} \ \pr^{\; q} \ = \ \left( {p+q} \atop p  \right) \ \pr^{\; p+q} 
\\
&& \prd \left( \pr^{\; p} \ \phi \right) \ = \ \Box \ \pr^{\; p-1} \ \f \ + \ 
\pr^{\; p} \ \prd \f \\
&& \prd \h^{\;k} \ = \ \pr \, \h^{\;k-1} \ , \label{etak}
\ea
hold. For instance, a special case of (\ref{etak}) is
\be
\prd \h^{\;2} \equiv \pr_1 \left( \h_{12} \,\h_{34} +  \h_{13} \, \h_{24} + 
\h_{14} \, \h_{23} \right) = \left( \pr_2 \, \h_{34} +  \pr_3 \, \h_{24} + 
\pr_4 \, \h_{23} \right) \equiv \pr \, \h \ ,
\ee
and the advantages of the compact notation should be evident.

The gauge transformation of the spin-$s$ field
then reads
\be
\delta  \phi \ = \ \pr \, \Lambda \ , \label{gaugepins}
\ee
while the generic spin-$s$ Fronsdal equation becomes
\be
{\cal F} \ = \ \Box \, \phi \  -\ \pr \ \prd \phi \ + \ \pr^{\;2}\, \phi\;' \ ,
\ee
whose gauge variation is
\be
\d {\cal F} \ = \ 3\ \pr^{\; 3} \L' \ .
\ee
The spin-$s$ Fronsdal operator satisfies in general the 
``anomalous'' Bianchi identity
\be
\prd {\cal F} \  - \ \frac{1}{2} \ \pr \, {\cal F}\;' \ = \ - \
\frac{3}{2} \ \pr^{\;3}\, \phi^{''} \ , \label{bianchifron}
\ee
where the difference with respect to the spin-3 case should be
noted,
and as a result one can define the Einstein-like tensor
\be
{\cal G} \ = \ {\cal F} \ - \ \frac{1}{2} \ \h \ {\cal F}\;' \ ,
\ee
such that
\be
\prd {\cal G} \ = \ -  \ \frac{3}{2} \ \pr^{\; 3}\, \phi^{''} \ - \ 
\frac{1}{2} \ \h \ \prd 
{\cal F}\;'  \ .
\ee
This relation is at the heart of the usual restrictions, present in the
Fronsdal formulation, to traceless gauge parameters and doubly traceless
fields, needed to ensure that
\be
\delta {\cal L} \ = \ \delta \phi \ {\cal G}  \label{deqspins}
\ee
vanish if $\d \f$ is given by eq. (\ref{gaugepins}).

One can now define recursively a sequence of kinetic operators, as
\be
{\cal F}^{(n+1)} \ = \ {\cal F}^{(n)} \ + \ \frac{1}{(n+1) (2 n + 1)} \ 
\frac{\pr^{\;2}}{\Box} \, {{\cal F}^{(n)}}\;' \ - \ \frac{1}{n+1} \ 
\frac{\pr}{\Box} \ 
\prd  {\cal F}^{(n)} \ , 
\ee
where ${\cal F}^{(1)}={\cal F}$, and an inductive argument then shows that
\be
\d {\cal F}^{(n)} \ = \ \left( 2 n + 1 \right) \ \frac{\pr^{\; 2 n + 1}}
{\Box^{\; n-1}} \ \Lambda^{[n]} \ ,
\ee
where, as anticipated, $\Lambda^{[n]}$ denotes the $n$-fold 
trace of the gauge parameter $\Lambda$.
Notice that this is only available for spin $s > 2 n + 1$, and therefore this
procedure yields a gauge-invariant kinetic operator after a
certain number of iterations. 

If, as in \cite{bargtod}, the gauge field $\phi_{1 \cdots s}$ 
is contracted with a vector $\xi$, it is simple to 
convince oneself that
traces and divergences of the resulting expression
\be
\hat{\Phi}(x,\xi) \ = \ \frac{1}{s!}\ \xi^{1} \cdots \xi^{s} \ \phi_{1 
\cdots s}
\ee
can be recovered applying to it the differential operators
$\pr_\xi \cdot \pr_\xi$ and $\pr_\xi \cdot \pr$, where $\pr_\xi$
denotes a derivative with respect to $\xi$. The least singular
non-local field equations obtained from the 
Fronsdal term
\be
\hat{\cal F}(\hat{\Phi}) = \left[ \ 
\Box \ - \ \xi \cdot \pr \ \ \prd \pr_\xi \ + \ (\xi \cdot \pr)^2 \ \pr_\xi \cdot \pr_\xi
  \ \right] 
\hat{\Phi} \ , \label{fronsdalxi}
\ee
 by successive iterations
can then be written in the compact form\\
\be
  \prod_{k=0}^{n-1} \left[\  1 \ +\  \frac{1}{(k+1)(2k+1)} \ \frac{(\xi \cdot \pr)^2}{\Box} \
 \pr_\xi \cdot \pr_\xi \ - \ \frac{1}{k+1} \ \frac{\xi \cdot \pr}{\Box} \ \pr_\xi \cdot \ \pr
 \  \right]  \ \hat{\cal F}(\hat{\Phi}) \ = \ 0 \  , \label{eqnonlocgen}
\ee\\
where for spin $s$ a fully gauge invariant 
operator is first obtained after $\left[ \frac{s+1}{2} \right]$ iterations.
Expanding this expression and combining it with its trace it is possible
to show that the field equations can all be reduced to the form
\be
{\cal F} \ =  \ \pr^{\; 3} \ {\cal H} \ , \label{fd3h}
\ee
where under a gauge transformation $\delta {\cal H} = 3 \ \L{\; '}$, and therefore
the local form (\ref{fronsdalxi}) can always be recovered from
(\ref{eqnonlocgen}) making use of the trace of the gauge parameter $\L$.

These kinetic operators satisfy the ``anomalous'' Bianchi identities
\be
\prd {\cal F}^{(n)} \ - \ \frac{1}{2n} \ \pr {{\cal F}^{(n)}}{\; '} \ = \ - \
\left( 1 + \frac{1}{2n}  \right) \ \frac{\pr^{\; 2n+1}}{\Box^{\; n-1}} \ 
\phi^{(n+1)} \ , \label{bianchin}
\ee
that generalize eq. (\ref{bianchi3}). This result can be also justified by
an inductive argument, and implies 
similar relations for successive traces of the ${\cal F}^{(n)}$,
\be
\prd {\cal F}^{(n)\, [k]} \ - \ \frac{1}{2(n-k)} \ 
\pr {{\cal F}^{(n)\, [k+1]}} \ = \
0 \ , \qquad ( k \leq n-1) \label{bianchink}
\ee
here written for $n$ large enough so that the ``anomaly'' on the
{\it r.h.s.} of (\ref{bianchin}) vanishes identically. Notice that
for odd spin $s=2n-1$ the second term vanishes for the last trace, so
that
\be
\prd {\cal F}^{(n)\, [n-1]} \ = \ 0 \ . \label{lastodd}
\ee

These generalized Bianchi identities suffice to define for all spin-$s$
fields fully gauge invariant analogues of the Einstein tensor,
\be
{\cal G}^{(n)} \ = \ \sum_{p \leq n} \ \frac{(-1)^p}{2^p \ p! \ 
\left( {n \atop p} \right)} \ \eta^p \ {\cal F}^{(n)\, [p]} \ 
\ee
that, for $n$ large enough, have vanishing divergence like their
spin-2 counterpart. This is attained directly by the subtractions for
all even spins, while for odd spins the last term
vanishes on account of (\ref{lastodd}).
From ${\cal G}^{(n)}$,  integrating eq. (\ref{deqspins}) one can then
construct generalized Lagrangians that are fully gauge invariant without any
restrictions on the gauge fields or on the gauge parameters.

\section{Geometric forms of the spin-$s$ field equations}

Following \cite{dewf}, one can define generalized connections 
of various orders in the derivatives for all spin-$s$
gauge fields. This can be done by an iterative procedure, so that,
in the compact notation of the previous section,
for any field of spin $s$ after $m$ iterations one can define
\be
\G^{(m)} \ = \ \frac{1}{m+1} \ \sum_{k=0}^{m}\ \fr{(-1)^{k}}{
\left(
{{m} \atop {k}}
\right)
}\ 
\pr^{\; m-k}\, \btd^{\; k} \, \f \ , \label{gammam}
\ee      
where we are now using two types of derivatives for two sets of
symmetrized indices, $\pr$ for the $s$ symmetric indices 
$(\b_1 \cdots \b_s)$ and $\btd$ for the other
$m$ symmetric ones $(\a_1 \cdots \a_m)$. It is simple to show, 
by an inductive argument, that the gauge transformation of $\G^{(m)}$ is
\be
\delta \, \G^{(m)} \ = \ \pr^{\; m+1} \, \L \ , \label{deltagammam}
\ee
where all $m$ indices of the first set are within the gauge parameter. 
Hence, 
\be
\G^{(s-1)} \ = \ \fr{1}{s} \ \sum_{k=0}^{s-1}\ \fr{(-1)^{k}}{
\left(
{{s-1} \atop {k}}
\right)
}\ 
\pr^{\; s-k-1}\, \btd^{\; k} \, \phi \ , \label{connection}
\ee
is the proper analogue of the Christoffel connection for
a spin-$s$ gauge field, since its gauge transformation 
contains a single term.
That these objects can be defined in general can
be also recognized noticing that the spin-$s$ gauge variation of 
eq. (\ref{gaugepins}) and the
rules of symmetric calculus of the previous section imply that
\be
\delta \left( \pr^{\; s-1} \ \phi \right) \ = \ s \ \pr^{\; s}\, \Lambda \ ,
\ee
and therefore one can in principle retrieve a composite connection
$\G_{\a_1 \cdots \a_{s-1}; \b_1 \cdots \b_s}$ such that
\be
\delta \; \G_{\a_1 \cdots \a_{s-1}; \b_1 \cdots \b_s}  \ = \ \pr_{\b_1} \cdots
\pr_{\b_s} \ \L^{ \a_1 \cdots \a_{s-1}} \label{gaugetransfgammas}
\ee
inverting the linear system
\be
\pr^{\; s-1} \ \phi \ = \ \Gamma_{\{s-1\};\{ s \} } \ ,
\ee
 with $\left( {2s-1 \atop s } \right)$ unknowns,
a higher-derivative analogue of the linearized metric 
postulate for Einstein gravity. 
Moreover, {\it all} $\G$'s with $m >s$ are gauge invariant, and
in particular 
\be
\G^{(s)} \ = \ \fr{1}{s+1}\ \sum_{k=0}^{s}\ \fr{(-1)^{k}}{
\left(
{{s} \atop {k}}
\right)\ 
}\ 
\pr^{\; s-k}\, \btd^{\; k} \, \phi \ , \label{riemann}
\ee      
is the proper analogue of the Riemann curvature
tensor. This generalized curvature
${\cal R}_{\alpha_1 \cdots \alpha_s;\beta_1 \cdots
\beta_s}$ is totally symmetric under the interchange of any two
indices within the two sets. In addition, as shown
in \cite{dewf},
\be
{\cal R}_{\alpha_1 \cdots \alpha_s;\beta_1 \cdots
\beta_s} \ = \ (-1)^s \ {\cal R}_{\beta_1 \cdots
\beta_s;\alpha_1 \cdots \alpha_s} \ ,
\ee
and a generalized cyclic identity holds. These concepts can also be
related to an interesting generalization of the exterior differential,
whereby the familiar condition $d^2=0$ is replaced by $d^{s+1}=0$ \cite{henn}.

There is another, perhaps more obvious way, to generate a gauge invariant
quantity from a connection $\G^{(s-1)}$ that transforms as in
(\ref{gaugetransfgammas}),
taking a curl with respect to any of its $\b$ indices. However,
the choice of \cite{dewf} has the virtue of simplicity, since it
results automatically in a tensor with two totally symmetric sets of indices.
If we now restrict our attention to the $\G$'s with $m$ {\it even},
and for the sake of clarity let $m=2n$, eq. (\ref{deltagammam}) 
implies that the total trace of $\G^{(2n)}$ over pairs of $\b$ indices,
$\G^{(2n) [n]}$, is in general a totally symmetric spin-$s$ tensor such that 
\be
\d \, \left( \, \frac{1}{\Box^{n-1}}\ \G^{(2n)\, [n]} \, \right) \  = \
\frac{\pr^{\; 2n+1}}{\Box^{\; n-1}} \ \L^{[n]} \ .
\ee
Up to an overall proportionality constant,
this is exactly the gauge transformation of our ${\cal F}^{(n)}$, the
corrected kinetic operators for spin-$s$ gauge fields, and in particular
if $s=2n$ $\G^{(2n)[n]}$ is gauge invariant and proportional to the spin-$s$
analogue of the Riemann tensor defined above. This therefore means that
the iterative procedure of the previous section is actually providing a r\^ole
for the higher-spin connections of \cite{dewf}, so that the 
geometric gauge-invariant equations for even spin $s=2n$ can be written
in the form
\be
\frac{1}{\Box^{n-1}} \ {\cal R}^{[n]}{}_{;\mu_1 \cdots \mu_{2n}} \ = \ 0 \ ,
\label{geomeven}
\ee
a natural generalization of the Einstein equation. 

The odd-spin case $s=2n+1$ presents a further minor subtlety, in that the
corresponding curvatures $\G^{(2n+1)}$ have an odd number of $\b$ indices.
The simplest option is in this case to take a trace over $n$ pairs of
$\b$ indices in $\G^{(2n+1)}$ and a divergence over the remaining one.
The end result for spin $s=2n+1$ is then
\be
\frac{1}{\Box^{n}} \ \prd {\cal R}^{[n]}{}_{;\mu_1 \cdots \mu_{2n+1}} \ = \ 0 \ ,
\label{geomodd}
\ee
in complete analogy with the Maxwell case. Notice that the Maxwell
and Einstein cases are the only ones when these geometric equations
are local, while the Fronsdal operators
provide local, albeit partly gauge fixed, forms for them.
As anticipated in the previous sections, these are the least singular
fully gauge invariant
kinetic operators, while more singular forms can be obtained combining 
eqs. (\ref{geomeven}) and (\ref{geomodd})
with their traces, as we saw in section 2.

\section{Fermionic equations}

One can also arrive at similar non-local geometric equations
for fermion fields. In this case the
local equations of \cite{ffron}
\be
{\cal S} \ \equiv \ i \, \left( {\not {\! \pr}} \, \psi - \pr \psisl \right) \ = \ 0
\ee
are gauge invariant under
\be
\delta \psi \ = \ \pr \, \e
\ee
only if the gauge parameter is subject to the constraint
\be
\esl \ = \ 0 \ . \label{fermiparf}
\ee

In addition, ${\cal S}$ satisfies the ``anomalous'' Bianchi identity
\be
\prd {\cal S} \ - \ \frac{1}{2} \, \pr \ {\cal S}{\; '} \ - \ 
\frac{1}{2} {\not {\! \pr}} \ssl \  
 = \ i \ \pr^{\; 2} \psisl\;' \ ,
\ee
and therefore the gauge variation of the generic Lagrangian
\be
\d {\cal L} \ = \ \d \bar{\psi} \left[\, {\cal S} \ - \ \frac{1}{2} \left(\, 
\eta \, {\cal S}' \ + \ \gamma \ssl \, \right) \, \right]
\ee
vanishes only if 
\be
\psisl\; ' \ = \ 0 \ , \label{fermiparf2}
\ee
the fermionic analogue of the double trace condition for boson fields.

It is convenient to notice that the fermionic operators 
for spin $s+1/2$ are related to the corresponding
bosonic operators for spin $s$ according to
\be
{\cal S}_{s+1/2} \ - \ \frac{1}{2} \, \frac{\pr}{\Box}\, {\not{\!\pr}} \, 
\ssl_{s+1/2} \ = \ i \ \frac{\not{\!\pr}}{\Box} \, {\cal F}_s(\psi) \ . 
\ee
This amusing link generalizes the obvious one between the Dirac and 
Klein-Gordon operators, and actually extends to their non-local 
counterparts. Hence, it allows one to relate corrected fermionic 
kinetic operators ${\cal S}^{(n)}$, defined recursively as
\be
{\cal S}^{(n+1)} \ = \ {\cal S}^{(n)} \ + \ \frac{1}{n(2n+1)} \,
\frac{\pr^{\; 2}}{\Box} \, 
{\cal S}^{(n)\; '} \ - \ \frac{2}{2n+1} \, \frac{\pr}{\Box} \, 
\prd {\cal S}^{(n)}
\ee
and such that
\be
\d \, {\cal S}^{(n)} \ = \ - \ 2 \, i \, n \ 
\frac{\pr^{\; 2n}}{\Box^{\; n-1}}\,  \esl^{\; [n-1]}
\ee
to the corresponding corrected bosonic operators of section 3, according to
\be
{\cal S}^{(n)}_{s+1/2} \ - \ \frac{1}{2n} \, \frac{\pr}{\Box}\, {\not{\!\pr}} 
\, 
\ssl_{s+1/2}^{(n)} \ = \ i \ \frac{\not{\!\pr}}{\Box} \, {\cal F}^{(n)}_s(\psi) \ . 
\ee

This relation also determines the ``anomalous'' Bianchi identities of the
${\cal S}^{(n)}$,
\be
\prd {\cal S}^{(n)} \ - \ \frac{1}{2n} \, \pr \ {\cal S}^{(n)\; '} 
\ - \ \frac{1}{2n}\,  
{\not {\! \pr}} \ssl^{(n)} 
\  = \ i \ \frac{\pr^{\; 2n}}{\Box^{\; n-1}} \psisl^{[n]} \ ,
\ee
and therefore the corrected Einstein-like operators
\be
{\cal G}^{(n)} \ = \ {\cal S}^{(n)} \ + \
\sum_{0 < p \leq n} \ \frac{(-1)^p}{2^p \ p! \ 
\left( {n \atop p} \right)} \ \eta^{p-1} \left[\  \eta \ 
{\cal S}^{(n)\, [p]} \ + \ \gamma  \ {\cal {\not {\! S}}}^{(n)\, [p-1]}
\ \right] \ .
\ee

The geometry underlying the bosonic case thus bears a close, if less
direct, relation to the fermionic operators ${\cal S}^{(n)}$, that
can also be retrieved from the iterated bosonic terms ${\cal F}^{(n)}$.  

 \vskip 24pt
\begin{flushleft} {\large \bf Acknowledgments}
\end{flushleft} 

It is a pleasure to thank M. Vasiliev for stimulating conversations.
This work was supported in part by
I.N.F.N., by the EC contract HPRN-CT-2000-00122, by the EC contract
HPRN-CT-2000-00148, by the INTAS contract 99-1-590 and by the MURST-COFIN
contract 2001-025492. D.F. was supported by an I.N.F.N. 
Pre-Doctoral Fellowship. This work was presented at the Third International
Sakharov Conference, Moscow, June 24-29, 2002, held at the Lebedev Institute,
that the authors would like to thank for the kind hospitality.

\end{document}